\documentclass[10pt,a4paper,twoside,dvipdfm]{article}
\usepackage{epsfig}
\usepackage{baltlat6}
\usepackage{array}
\usepackage{wrapfig}
\usepackage{dcolumn}
\begin{document}
\ \
\vspace{0.5mm}
\setcounter{page}{1}
\vspace{8mm}

\titlehead{Baltic Astronomy, vol.\,18, 1--18, 2009}

\titleb{YOUNG STARS IN THE CAMELOPARDALIS DUST AND\\ MOLECULAR CLOUDS.
V. MORE YSOs CONFIRMED\\ SPECTROSCOPICALLY}

\begin{authorl}
\authorb{C. J. Corbally}{1} and
\authorb{V. Strai\v{z}ys}{2}
\end{authorl}

\begin{addressl}
\addressb{1}{Vatican Observatory Research Group, Steward Observatory,
Tucson, Arizona 85721, U.S.A.; corbally@as.arizona.edu}

\addressb{2}{Institute of Theoretical Physics and Astronomy, Vilnius
University,\\  Go\v{s}tauto 12, Vilnius LT-01108, Lithuania;
straizys@itpa.lt}
\end{addressl}

\submitb{Received 2009 May 15; accepted 2009 May 30}

\begin{summary} Far red spectra for 22 stars in the Camelopardalis and
the northern Perseus dark clouds, suspected to be pre-main-sequence
objects (YSOs), are obtained.  This evolutionary status is confirmed for
ten stars located in the dust and molecular cloud close to the high-mass
protostar GL\,490, four stars near the H\,II region Sh2-205 and one star
in the dark cloud TGU\,1041.  All of these objects exhibit emission in
the H$\alpha$ line and some of them emission in the O\,I and Ca\,II
lines.  The spectral energy distributions, equivalent widths of the
emission lines and approximate spectral classes are determined.
Evolutionary stages of the stars are estimated from 2MASS {\it J, H,
K}$_s$, IRAS and MSX infrared photometry.  Now we have spectral
confirmation of the YSO status for 14 stars in the GL\,490 area and 8
stars at Sh2-205.  Their spectral types are from A to K, but most of
them are either Herbig Ae stars or intermediate objects between T Tauri
type and Herbig stars.  Both these star forming regions are located near
the outer edge of the Local arm at a distance of $\sim$\,900 pc.
\end{summary}

\begin{keywords} stars:  pre-main sequence -- stars: emission-line --
star-forming regions: individual (GL\,490, Sh2-205)
\end{keywords}

\resthead{Young stars in the Camelopardalis dust and molecular clouds.
V.}{C. J. Corbally, V. Strai\v zys}

\sectionb{1}{INTRODUCTION}

In the direction of Camelopardalis our line of sight crosses the Local,
Perseus and Outer arms.  In the two nearest arms numerous dust and
molecular clouds with continuing star formation are present, see the
review in Strai\v{z}ys \& Laugalys (2008b).  In the first three papers
of this series (Strai\v{z}ys \& Laugalys 2007a,b, 2008a, Papers I, II
and III) about 200 objects, applying the 2MASS, IRAS and MSX photometry
data, were suspected as being pre-main-sequence stars in different
stages of evolution.  In Paper IV (Corbally \& Strai\v{z}ys 2008) the
young stellar object (YSO) status was confirmed for 15 brightest stars
in this sample -- their far-red spectra in the 600--950 nm range exhibit
emission lines in H$\alpha$, O\,I, Ca\,II and P9.

About 50 of the suspected YSOs are concentrated in the
3\degr\,$\times$\,3\degr\ area with the center at $\ell$, $b$\,=
142.5\degr, +1.0\degr, in the vicinity of the massive protostar GL\,490,
in the densest part of the dust cloud TGU\,942 (Dobashi et al. 2005)
located at the outer edge of the Local arm at a distance of $\sim$\,900
pc.  We decided to verify the evolutionary status of 13 additional stars
in this region by new far-red spectral observations.  Additionally, we
included in the program eight stars in the vicinity of the H\,II region
Sh2-205 located near the Camelopardalis and Perseus border at $\ell$,
$b$ = 148\degr, --1\degr\ and one star from the catalog of Paper II.
The list of the observed stars for which emission-lines were found is
presented in Table 1. The stars are designated in the chart with
Galactic coordinates in Figure 1, and their identification charts of a
1.8\arcmin\,$\times$\,1.8\arcmin\ size are given in Figure 2.

\begin{table}[!t]
\begin{center}
\vbox{\small\tabcolsep=4pt
\parbox[c]{120mm}{\baselineskip=10pt
{\normbf\ \ Table 1.}{\norm\ List of the investigated stars in which
emission lines were found. $V$ is the green and
$F$ is the red photographic magnitudes taken from GSC\,2.3.2 (Lasker et
al. 2008). In the last
column, YSO means a suspected young stellar object, H$\alpha$
means a star with H$\alpha$ emission found in objective-prism spectra.
\lstrut}}
\begin{tabular}{lccccccl}
\tablerule
 Name  &  RA\,(2000) & DEC\,(2000) & $\ell$ & $b$~~~   &     $V$    &  $F$~~   &Type \\
\tablerule
SL\,176   & 3 17 24.4 &  +57 54 16 &   141.353&   +0.351 &  14.85  &  14.49 &  YSO\\[2pt]
SL\,175   & 3 20 53.3 &  +58 49 39 &   141.245&   +1.375 &  13.95  &  13.43 &  YSO\\[2pt]
SL\,153   & 3 24 27.7 &  +57 37 08 &   142.299&   +0.619 &  18.22  &  16.51 &  YSO\\[2pt]
SL\,144   & 3 26 20.6 &  +58 42 41 &   141.900&   +1.666 &  17.65  &  16.30 &  YSO\\[2pt]
SL\,177   & 3 27 58.7 &  +58 58 35 &   141.927&   +2.004 &  17.41  &  16.38 &  YSO\\[2pt]
SL\,183   & 3 28 47.7 &  +57 55 54 &   142.604&   +1.201 &  14.50  &  14.02 &  YSO\\[2pt]
SL\,184   & 3 30 03.2 &  +58 05 34 &   142.650&   +1.428 &  18.53  &  17.28 &  YSO\\[2pt]
SL\,158   & 3 30 05.6 &  +58 13 26 &   142.580&   +1.539 &  15.34  &  14.63 &  YSO\\[2pt]
SL\,163   & 3 32 53.3 &  +58 27 52 &   142.743&   +1.946 &  18.44  &  17.51 &  YSO\\[2pt]
SL\,165   & 3 34 00.7 &  +58 16 36 &   142.972&   +1.878 &  15.25  &  15.40 &  YSO\\[2pt]
Gahm 4    & 3 45 12.8 &  +52 14 38 &   147.849&  --2.014 &  13.52  &  12.76 &  H$\alpha$\\[2pt]
Gahm 11   & 3 56 14.1 &  +52 26 03 &   149.041&  --0.809 &  13.02  &  12.40 &  H$\alpha$\\[2pt]
Gahm 24   & 3 56 55.2 &  +52 51 20 &   148.849&  --0.420 &  15.08  &  14.43 &  H$\alpha$\\[2pt]
Gahm 2    & 3 58 13.8 &  +52 43 11 &   149.088&  --0.396 &  14.11  &  13.39 &  H$\alpha$ \\
SL\,136   & 4 31 52.5 &  +49 04 44 &   155.432&   +0.635 &  15.41  &  14.72 &  YSO\\
\tablerule
\end{tabular}
}
\end{center}
\vskip-3mm
\end{table}


\begin{table}[!th]
\begin{center}
\vbox{\footnotesize\tabcolsep=2pt
\parbox[c]{110mm}{\baselineskip=10pt
{\normbf\ \ Table 2.}{\norm\ Equivalent widths of emission lines and spectral
classification.}}
\begin{tabular}[t]{lrD..{-1}D..{-1}D..{-1}D..{-1}D..{-1}D..{-1}l}
\tablerule
\multicolumn{1}{c}{Star}  &
\multicolumn{1}{c}{Obs. date} &
\multicolumn{1}{c}{EW\,6563} &
\multicolumn{1}{c}{EW\,8446} &
\multicolumn{1}{c}{EW\,8498} &
\multicolumn{1}{c}{EW\,8542} &
\multicolumn{1}{c}{EW\,8662}&
\multicolumn{1}{c}{EW\,9226} &
\multicolumn{1}{c}{Spectral}  \\
      &  & \multicolumn{1}{c}{H$\alpha$} &
\multicolumn{1}{c}{O\,I} &
\multicolumn{1}{c}{Ca\,II} &
\multicolumn{1}{c}{Ca\,II} &
\multicolumn{1}{c}{Ca\,II} &
\multicolumn{1}{c}{P9} &
\multicolumn{1}{c}{class} \\
\tablerule
SL\,176   & 2008-10-19 & -8.3  &       &       &       &       &         & G0e\,(1) \\[2pt]
SL\,175   & 2008-10-19 & -3.1  &       &       &       &       &         & F0e\,(0.5) \\[2pt]
SL\,153   & 2008-10-19 & -43.4 &       & -1.6  & -1.0  &       &         & F0e\,(2) \\[2pt]
SL\,144   & 2008-10-19 & -14.5 &       & -2.0  & -1.8  &       &         & G0e\,(1)   \\[2pt]
SL\,177   & 2008-10-19 & -45.7 & -5.0  & -7.2  & -12.7 & -6.4  &         & A2e\,(3) \\[2pt]
SL\,183   & 2008-10-19 & -21.7 &       &       &       &       &         & A2e\,(2)   \\[2pt]
SL\,184*  & 2008-10-21 & -24.3 &       &       &       &       &         & F:\,e\,(0.5) \\[2pt]
SL\,158   & 2008-10-19 & -17.1 &       &       &       &       &         & A0e\,(2)    \\[2pt]
SL\,163*  & 2008-10-21 & -17.4 &       &       &       &       &         & F:\,e\,(0.5) \\[2pt]
SL\,165   & 2008-10-20 & -31.9 & -2.0  & -6.7  & -6.5  & -4.8  &         & G5e\,(2) \\[2pt]
Gahm 4    & 2008-10-20 & -38.7 & -2.9  &       &       &       &         & F0e\,(3) \\[2pt]
Gahm 11*  & 2008-10-20 & -26.0 & -2.6  &       &       &       &         & F0e\,(2) \\[2pt]
Gahm 24*  & 2008-10-20 & -34.5 & -3.1  &       &       &       &         & A5e\,(2) \\[2pt]
Gahm 2*   & 2008-10-20 & -61.2 & -4.4  &       &       &       &  -2.3   & F0e\,(3)  \\[2pt]
SL\,136   & 2008-10-21 & -39.3 & -4.5  &       &       &       &         & G0e\,(3)  \\
\noalign{\vskip1mm}
RY Tau    & 2008-10-19 & -11.7 & -1.0 & -2.6 & -1.0 & -1.0 &   & G0e\,(2)          \\[2pt]
CW Tau    & 2008-10-19 & -243.5& -8.1 & -30.0 & -24.9 & -25.0& -4.8   &  K5e\,(3)  \\[2pt]
DF Tau    & 2008-10-19 & -103.1 &       &       &       &       &         &  M0e\,(4) \\[2pt]
DG Tau    & 2008-10-21 & -79.4 & -2.9 & -30.4 & -20.1  & -28.8 & -3.6   &  G:\,e\,(5) \\[2pt]
DH Tau    & 2008-10-21 & -1.1  &      &       &        &       &        &  M0e
\\
\tablerule
\end{tabular}
}
\end{center}

\noindent $^*$~~{\footnotesize
SL\,163 and SL\,184: the spectra are very noisy; Gahm 11, 24 and 2: slight nebular emission.}
\vskip-2mm
\end{table}

\sectionb{2}{SPECTRAL OBSERVATIONS}

The spectra were taken on 2008 October 19--21 with the Boller \& Chivens
spectrograph on the Steward Observatory 2.3 m telescope at Kitt Peak,
using the 400 g/mm red-blazed grating, giving a resolution of 5.7 \AA\
and a range from 6070 to 9390 \AA\ on the BCSpec 1200\,$\times$\,800 CCD
detector.  The slit width was 1.5\arcsec.

The spectra, shown in Figure 3 in a widened form, were reduced using
IRAF software.  \footnote{~IRAF is distributed by the National Optical
Astronomy Observatory, which is operated by the Association of
Universities for Research in Astronomy (AURA) under cooperative
agreement with the National Science Foundation.} An additional step to
the reduction procedures followed in Paper IV was to make a far-red
fringing correction from blank sky spectra.  This had the effect of
reducing the H$_{2}$O and O$_{2}$ telluric band strengths as well as the
fringing for the low {\it S/N} spectra.  Detector glitches near 7170 and
8180
\AA\ were also much reduced by this step.  The energy distributions,
shown in Figure 4 (a--o), were obtained by calibrating the spectra with
the spectrophotometric standard Hiltner 600 and removing an atmospheric
extinction curve typical for Kitt Peak.  For the classification of stars
the criteria described by Danks \& Dennefeld (1994) in the far-red and
near-infrared spectral region were applied.  The results of
classification are only approximate since the spectral features at this
resolution and {\it S/N} do not give the discrimination of luminosities
and, in some cases, even of spectral subclasses.  The emission line
intensities of the spectral classifications are scaled in relation to
those given by Herbig (1962, Table I) for a few T Tauri stars whose
spectra were observed in the same nights as the program stars.

The equivalent widths of the prominent emission features (H$\alpha$,
Ca\,II triplet, O\,I at 8498 \AA\ and P9 at 9226 \AA) were measured with
the IRAF `splot' utility.  They are taken across the whole line,
including any absorption wings, so positive EWs will result when there
is emission in just the line cores.  Since these EW measures rely on a
by-eye estimate of the continuum level, three such measures were taken
and their mean values are given in Table 2. EWs were measured also in
the spectra of five T Tauri stars, RY Tau, CW Tau, DF Tau, DG Tau and DH
Tau, having very different emission-line intensities.

\newpage

\sectionb{3}{DESCRIPTION OF INDIVIDUAL OBJECTS}

In this section we give more information about the investigated
objects and discuss the results of their spectral classification,
emission line intensities and spectral energy distribution.

\vskip1mm

{\bf SL\,176 = 2MASS J03172447+5754136 = IRAS\,03135+5743}

The object is located at the lower edge of the dust cloud surrounding
GL\,940.  The star can be identified with IRAS\,03135+5743 which has
reliable flux only in the 25 $\mu$m band.  This point shows that the SED
curve in the log\,$\lambda$$F_{\lambda}$ vs log\,$\lambda$ coordinates
at $\lambda$\,$>$\,2 $\mu$m shows a slow decline typical of YSOs of
class II (T Tauri stars).  Spectral type (G0e) and the equivalent width
of H$\alpha$ emission (--8.3 \AA) given in Table 2 are in agreement with
a T Tauri star having a relatively moderate envelope or ring.  The
emission in lines of O\,I and Ca\,II at our resolving power is not
detectable.  In the $J$--$H$ vs.\,$H$--$K_s$ diagram the star lies only
$\sim$\,0.1 mag below the intrinsic T Tauri star line (Meyer et al.
1997).

\vskip1mm

{\bf SL\,175 = 2MASS J03205334+5849395}

The object is located in a relatively transparent direction, in the
outskirts of the GL\,490 cloud, only 101\arcsec\ from the star
HDE\,237121 of spectral type B0.5\,V, a member of the Cam OB1
association.  In the $J$--$H$ vs.\,$H$--$K_s$ diagram SL\,175 lies about
0.3 mag below the intrinsic T Tauri star line.  Our spectrum (F0e) and
the faint emission of H$\alpha$ ({\it EW}\,=\,--3.1 \AA) are consistent
with a post T Tauri or pre-Herbig star on the horizontal radiative track
from the T Tauri region to the main sequence (YSO of class III).

\vskip1mm

{\bf SL\,153 = 2MASS J03242757+5737064}

The object is located in the direction of a local dust cloud density
enhancement.  The star is not present in the IRAS and MSX databases,
consequently, we do not know if it has an excess of infrared radiation
beyond 2 $\mu$m.  However, its spectrum exhibits very strong H$\alpha$
emission ({\it EW}\,=\,--43.4 \AA) and much fainter emissions in the
Ca\,II triplet.  Its spectral type is F0e, and it again may be a
pre-Herbig Ae/Be star.

{\bf SL\,144 = 2MASS J03262053+5842412}

The object is located in the dust condensation, only 12\arcmin\ from
GL\,490. In the $J$--$H$ vs.\,$H$--$K_s$ diagram the star lies about 0.2
mag above the intrinsic T Tauri star line. Since the star is absent in
the IRAS and MSX databases, we cannot know about the presence of its
infrared excess. However, since its spectrum exhibits quite strong
H$\alpha$ line ({\it EW}\,=\,--14.5 \AA) and Ca\,II lines, the star is a
good candidate for T Tauri stars of spectral class G0e.

\vskip1mm

{\bf SL\,177 = 2MASS J03275850+5858341}

The object is located about 11\arcmin\ north of GL\,490, only 1\arcmin\
from 2MASS J03275562\\+5859269 = IRAS 03239+5849, an infrared and radio
source.  We classify SL\,177 as an A2e star with strong emissions in
H$\alpha$, O\,I and Ca\,II, and this is consistent with a Herbig Ae
object having a dense disk or envelope.  The H$\alpha$ emission for this
star was recently confirmed by the IPHAS survey (Witham et al. 2008;
Gonz\'alez-Solares et al. 2008).

\vskip1mm

{\bf SL\,183 = 2MASS J03284786+5755560 = IRAS 03248+5745}

The object is located in one of the dust condensations, 0.9\degr\ from
GL\,490. In the $J$--$H$ vs.\,$H$--$K_s$ diagram the star lies about 0.3
mag below the intrinsic T Tauri star line, therefore, in Paper III it
was suspected to be a Herbig Ae/Be star. The two reliable IRAS points at
12 and 25 $\mu$m give much stronger intensity of the star than in the
{\it JHK}$_s$ passbands, consequently, the star is an YSO of class I. In
our spectrum the {\it EW} of H$\alpha$ is --21.7 \AA, and the star is
of A2e spectral type.

\vskip1mm


\begin{figure}[!t]
\centerline{\psfig{figure=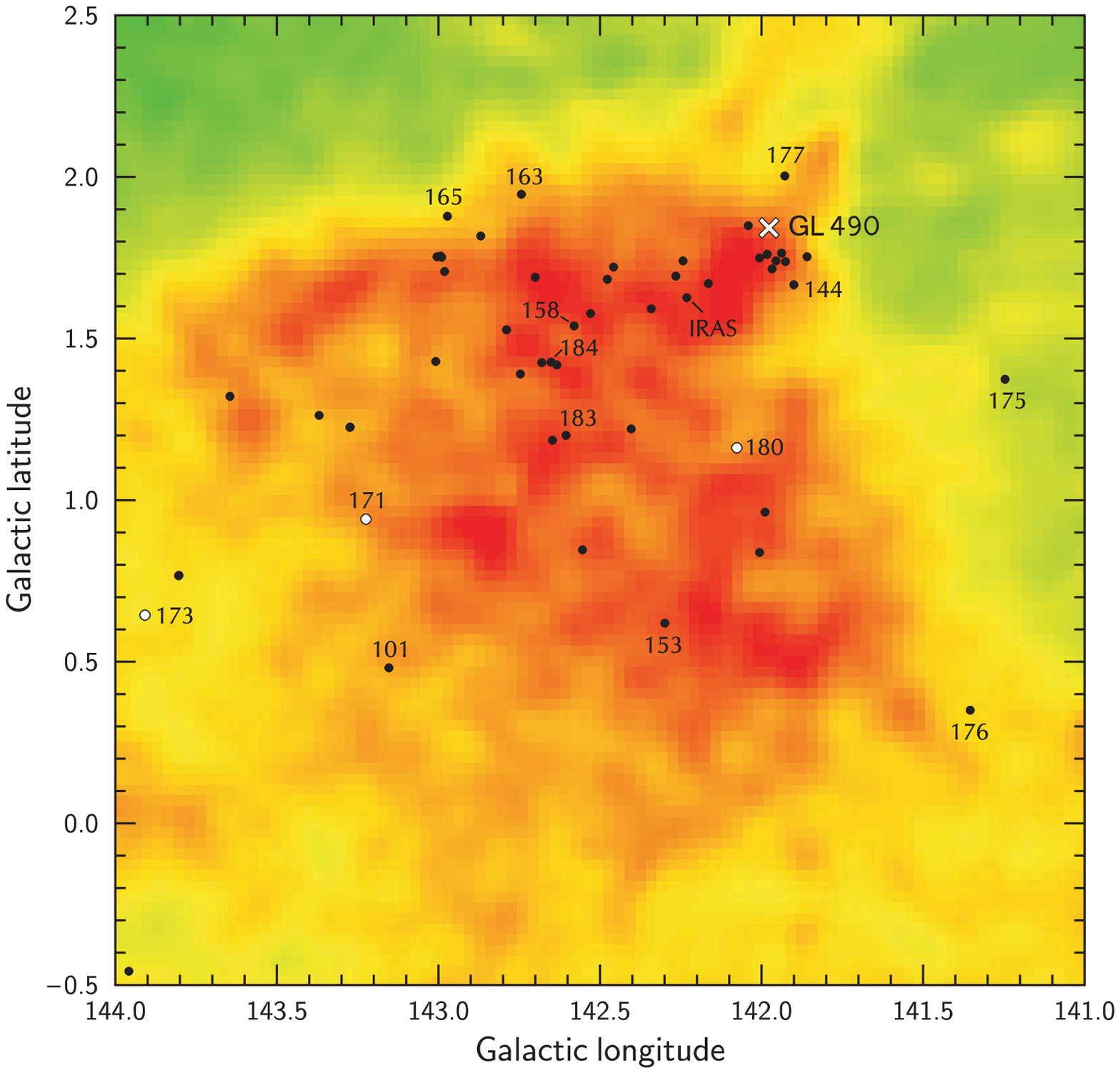,width=105mm,angle=0,clip=true}}
\vspace{0mm}
\captionb{1}{Positions of the suspected YSOs in the GL\,490 area
in the Galactic coordinates. Dust clouds from the Dobashi et al. (2005)
atlas are shown in the background. For the stars
analyzed in Paper IV and this paper, SL numbers are given. The white
circles designate three stars that have no emission lines. The star
marked as `IRAS' is IRAS 03243+5819.
}
\end{figure}

{\bf SL\,184 = 2MASS J03300294+5805348 = IRAS 03260+5755}

The object is located in a dust condensation, 0.8\degr\ from GL\,490.
In the $J$--$H$ vs.\,$H$--$K_s$ diagram SL\,184 lies about 0.3 mag above
the intrinsic T Tauri star line.  The star has only one reliable
measurement in the IRAS passbands, at 25 $\mu$m, in which its intensity
is comparable with that in $J$, $H$ and $K_s$.  Thus, the star belongs
to the YSO class II.  Our spectrum of this star is very noisy, but it
shows a strong H$\alpha$ with {\it EW} about --24.3 \AA\ and spectral
type F:\,e.  All the data listed favor the star to be of intermediate
type between T Tauri and Ae/Be stars.  The presence of emission in
H$\alpha$ was confirmed by the IPHAS survey (Witham et al. 2008;
Gonz\'alez-Solares et al. 2008).

\vskip1mm

{\bf SL\,158 = 2MASS J03300545+5813253 = IRAS 03261+5803}

The spectrum of this star was first shown in Paper IV.  In the $J$--$H$
vs.\,$H$--$K_s$ ~diagram SL\,158 ~lies about 0.15 mag above the
~intrinsic
T Tauri star line.  ~The star has reliable photometry in the IRAS 12 and
25 $\mu$m passbands and the MSX


\vbox{
\centerline{\psfig{figure=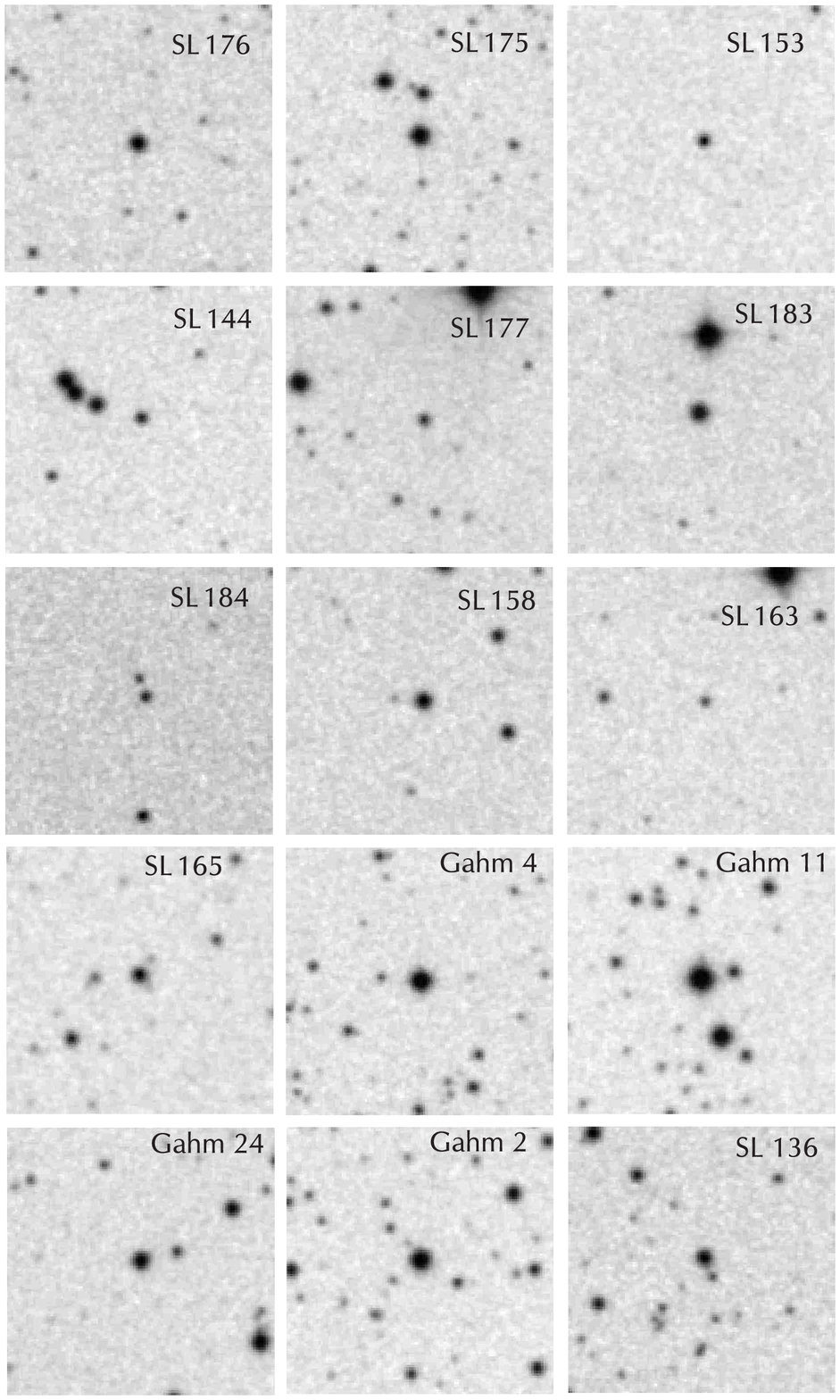,width=110mm,angle=0,clip=true}}
\vspace{0mm}
\captionb{2}{Identification charts of emission-line stars.  The fields
of
1.8\arcmin\,$\times$\,1.8\arcmin\ size are DSS2 red copies taken from
the Internet's Virtual Telescope SkyView.}
}

\newpage


\vbox{
\centerline{\psfig{figure=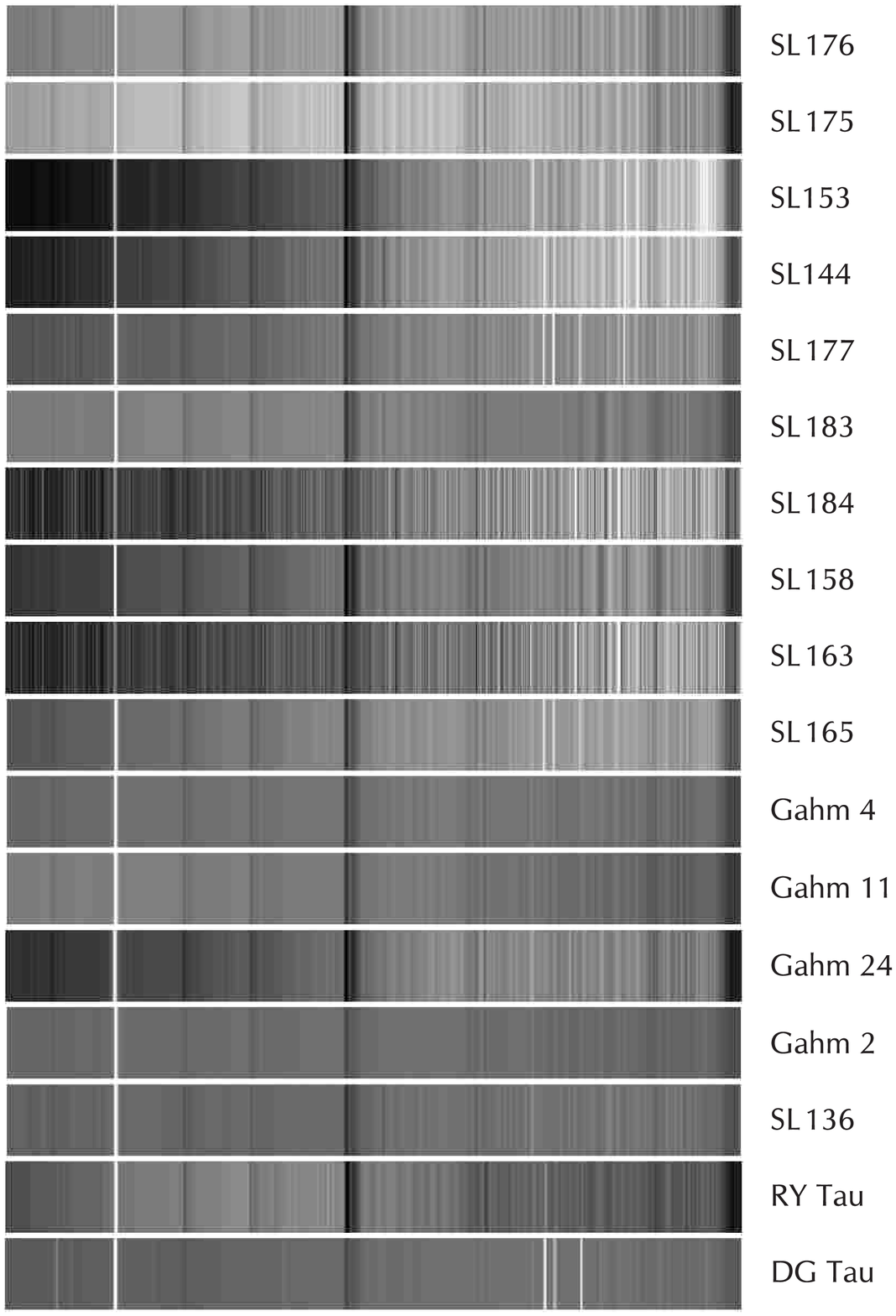,width=124mm,angle=0,clip=true}}
\vspace{2mm}
\captionb{3}{The widened spectra of the investigated emission-line
stars. For comparison, the spectra of two T Tauri type stars are given
at the bottom. The telluric H$_2$O and O$_2$ bands have not
been excluded.}
}
\newpage


\vbox{\begin{center}
\psfig{figure=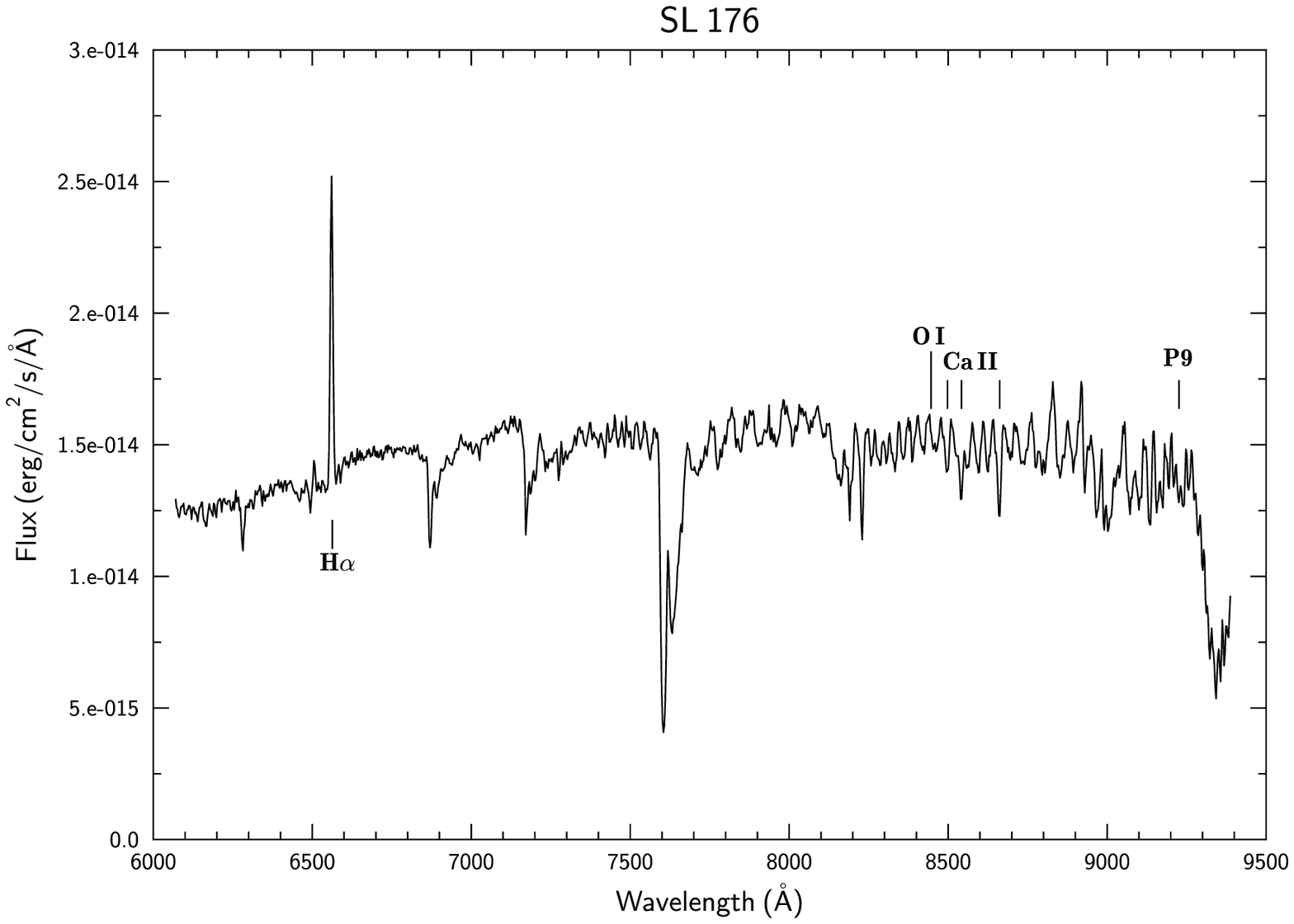,width=124mm,angle=0,clip=true}
\vskip2mm
\psfig{figure=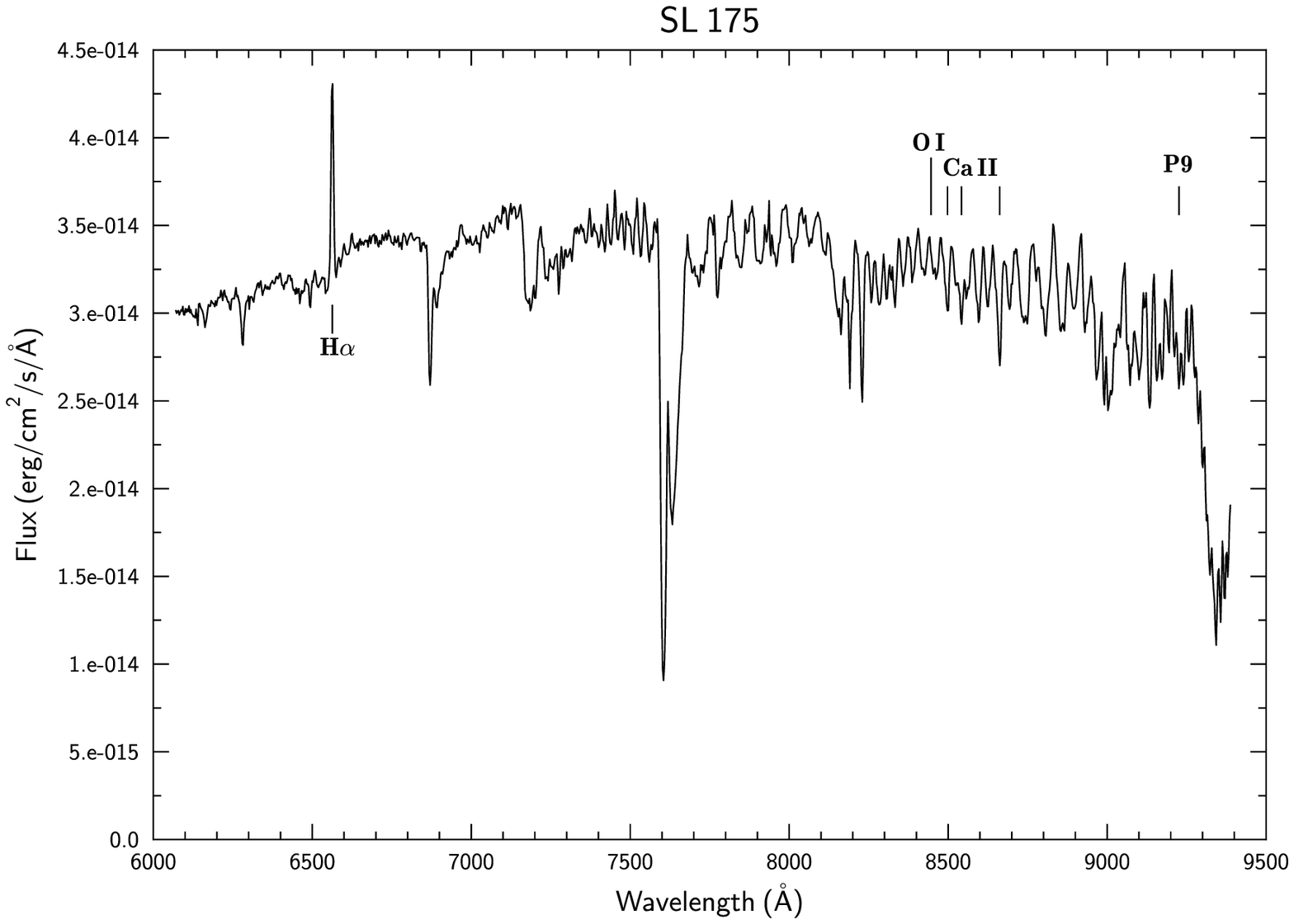,width=124mm,angle=0,clip=true}
\vskip.5mm
\captionc{4a and 4b}{Spectral energy distributions for the stars
SL\,176 and SL\,175.
}
\end{center}
}

\vbox{\begin{center}
{\psfig{figure=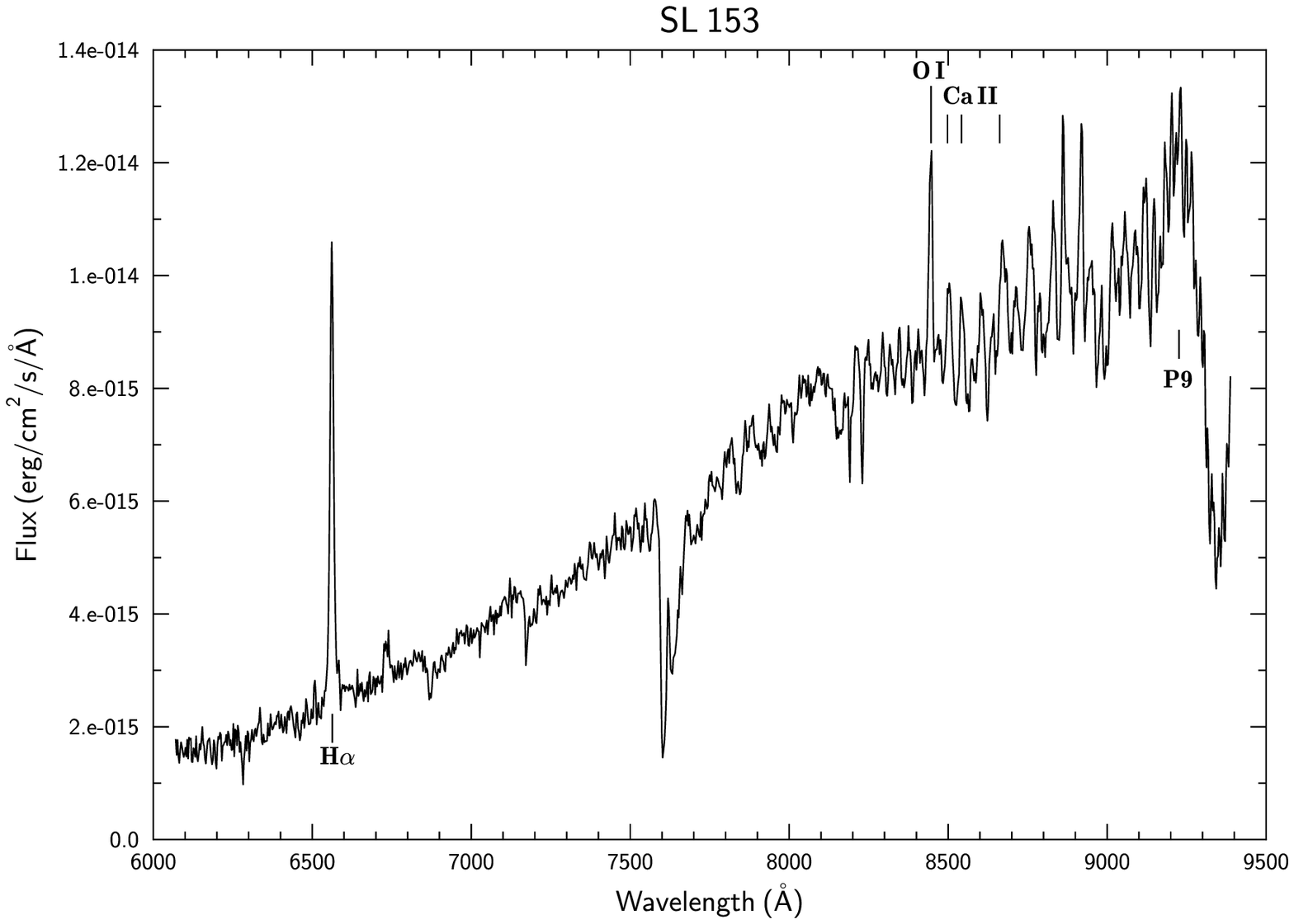,width=124mm,angle=0,clip=true}}
\vskip2mm
\psfig{figure=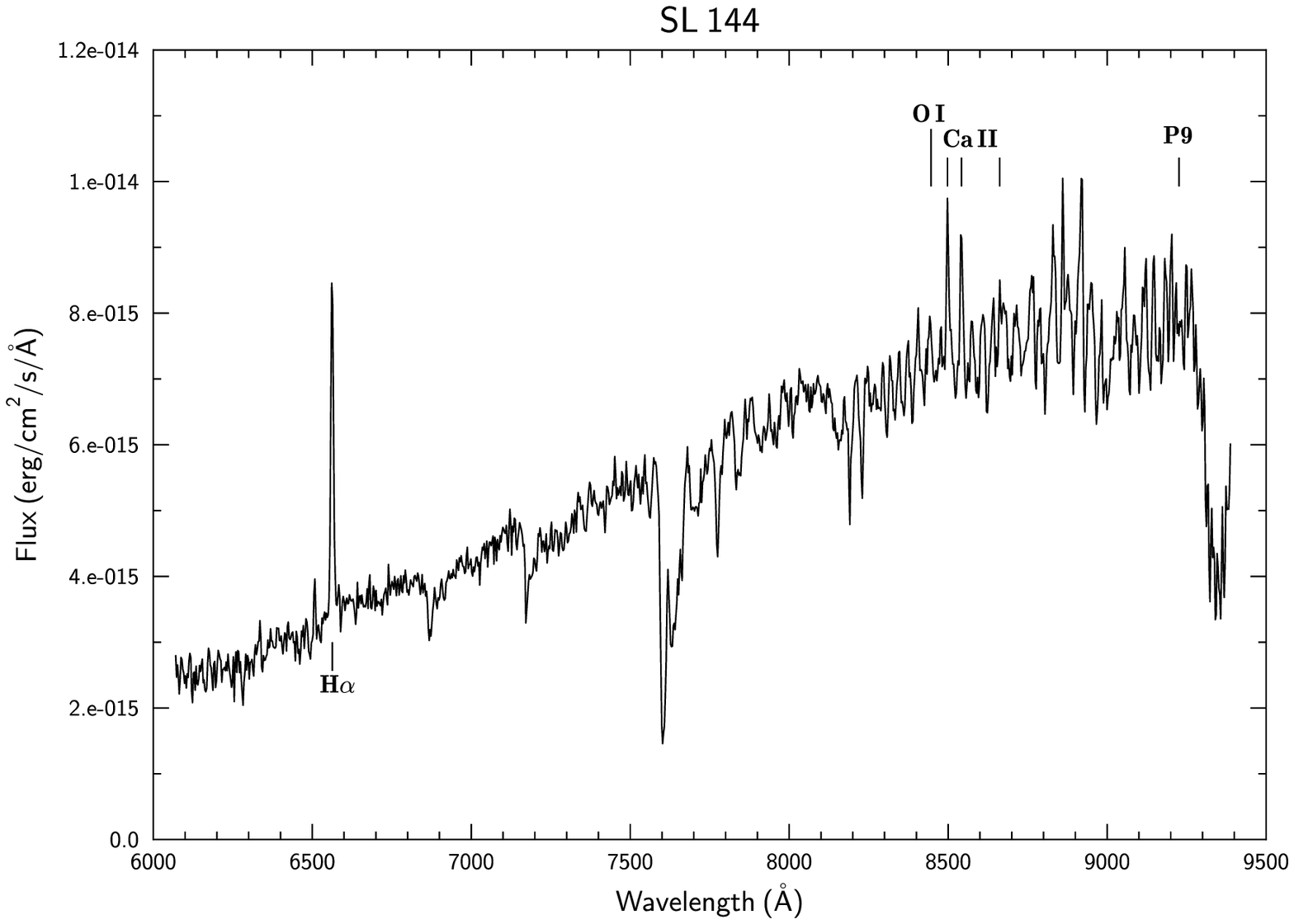,width=124mm,angle=0,clip=true}
\vskip.5mm
\captionc{4c and 4d}{Spectral energy distributions for the stars
SL\,153 and SL\,144.}
\end{center}
}

\vbox{\begin{center}
{\psfig{figure=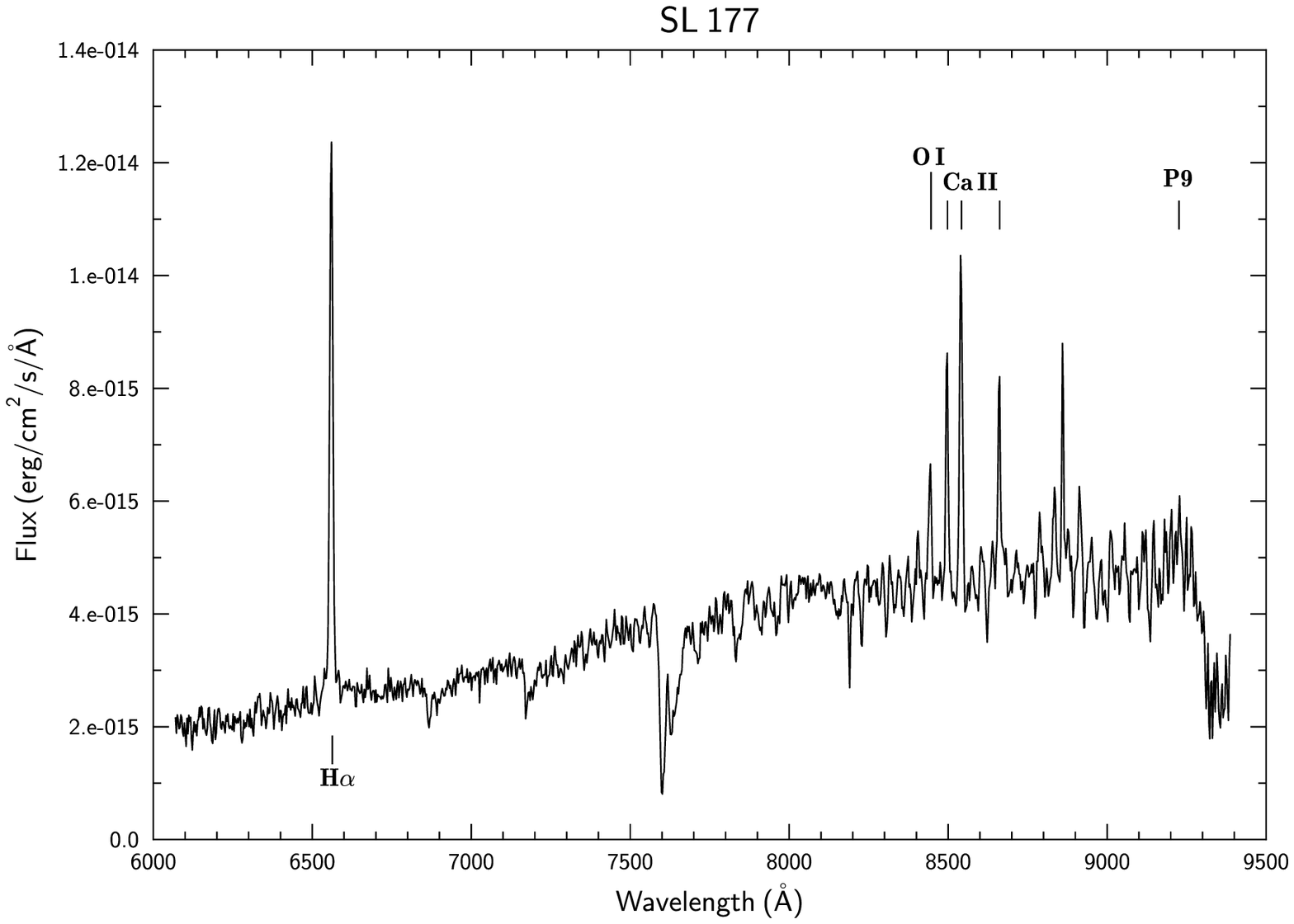,width=124mm,angle=0,clip=true}}
\vskip2mm
\psfig{figure=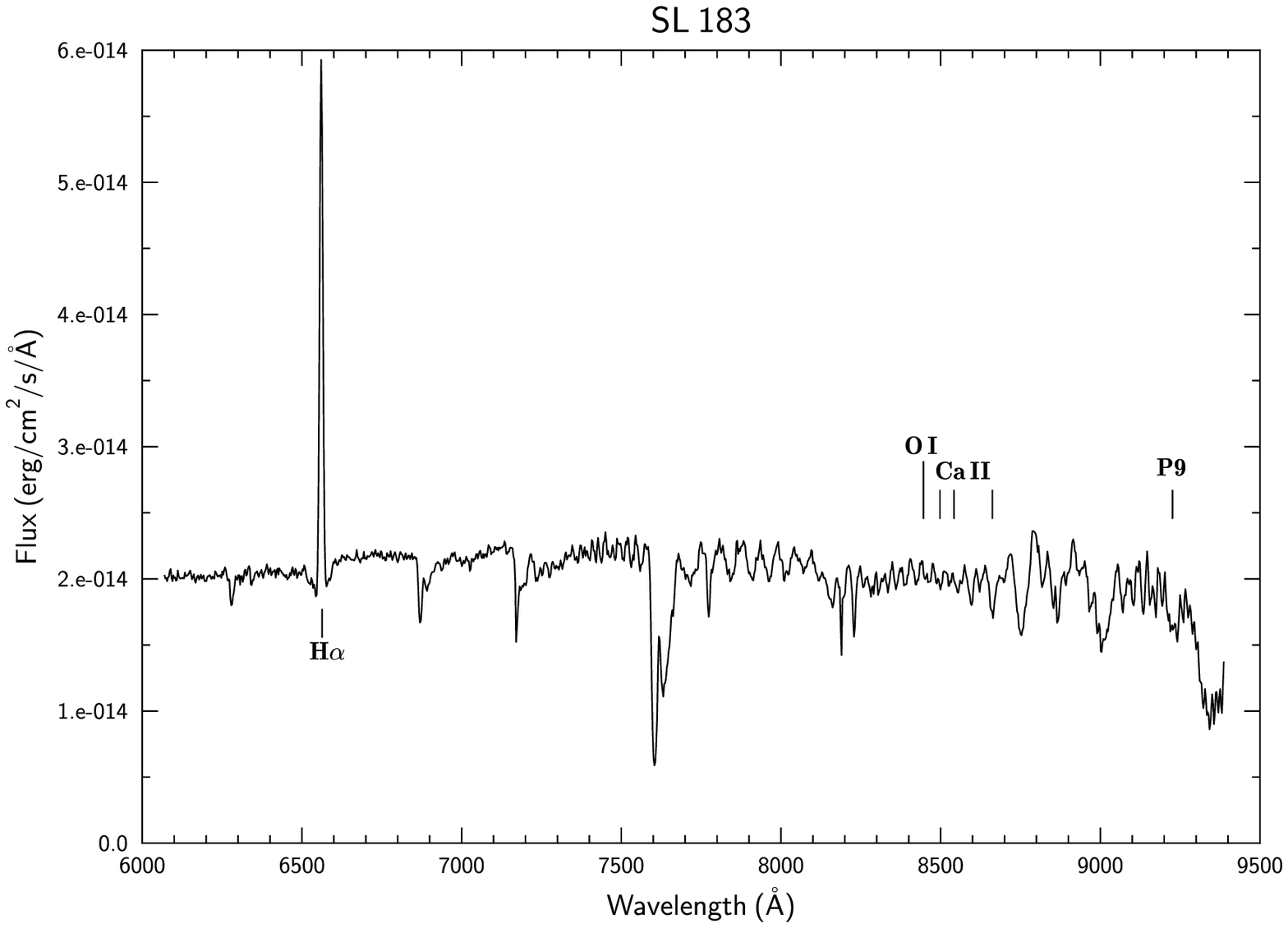,width=124mm,angle=0,clip=true}
\vskip.5mm
\captionc{4e~and~4f}{Spectral energy distributions for the stars
SL\,177 and SL\,183.}
\end{center}
}

\vbox{\begin{center}
{\psfig{figure=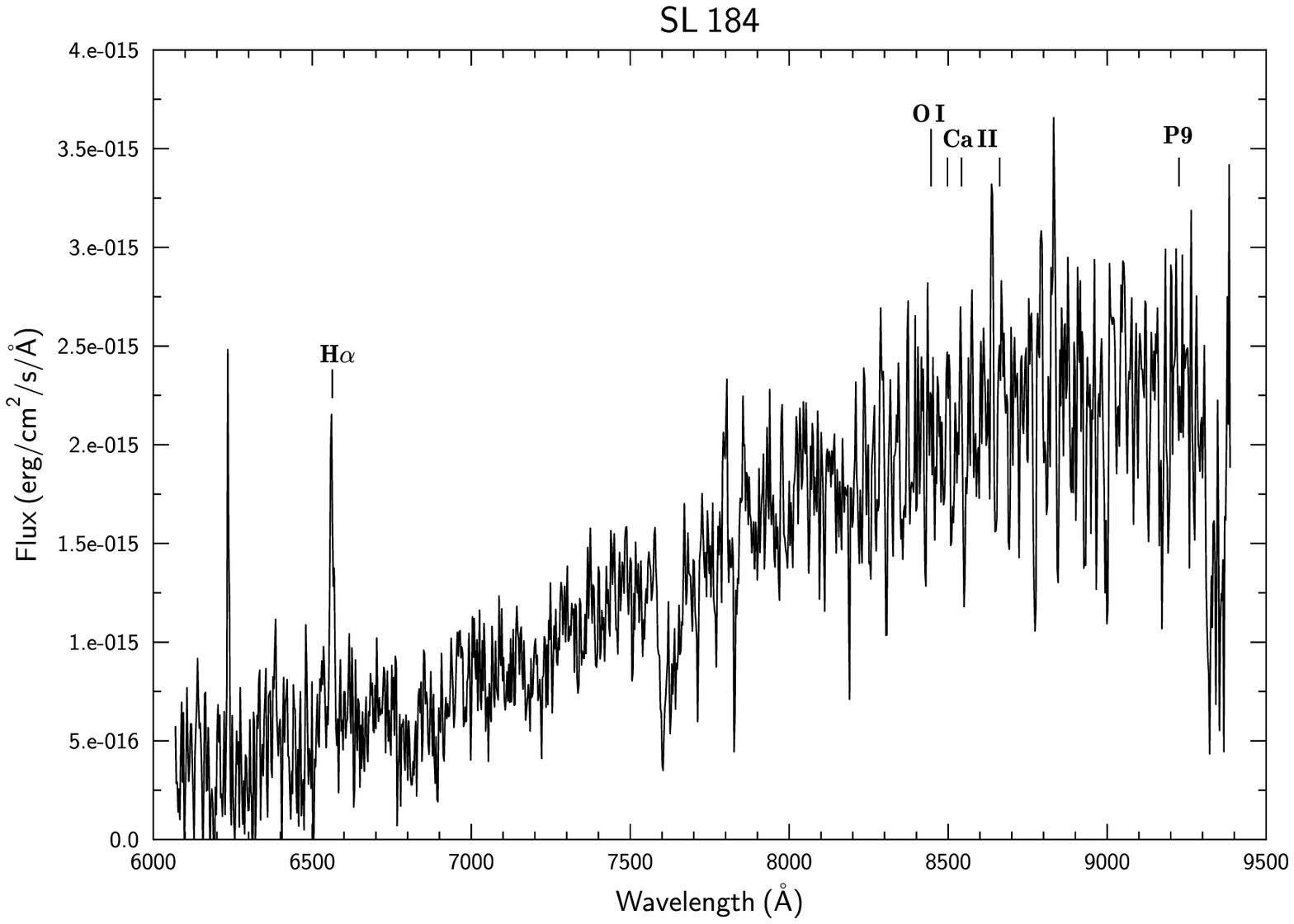,width=124mm,angle=0,clip=true}}
\vskip2mm
\psfig{figure=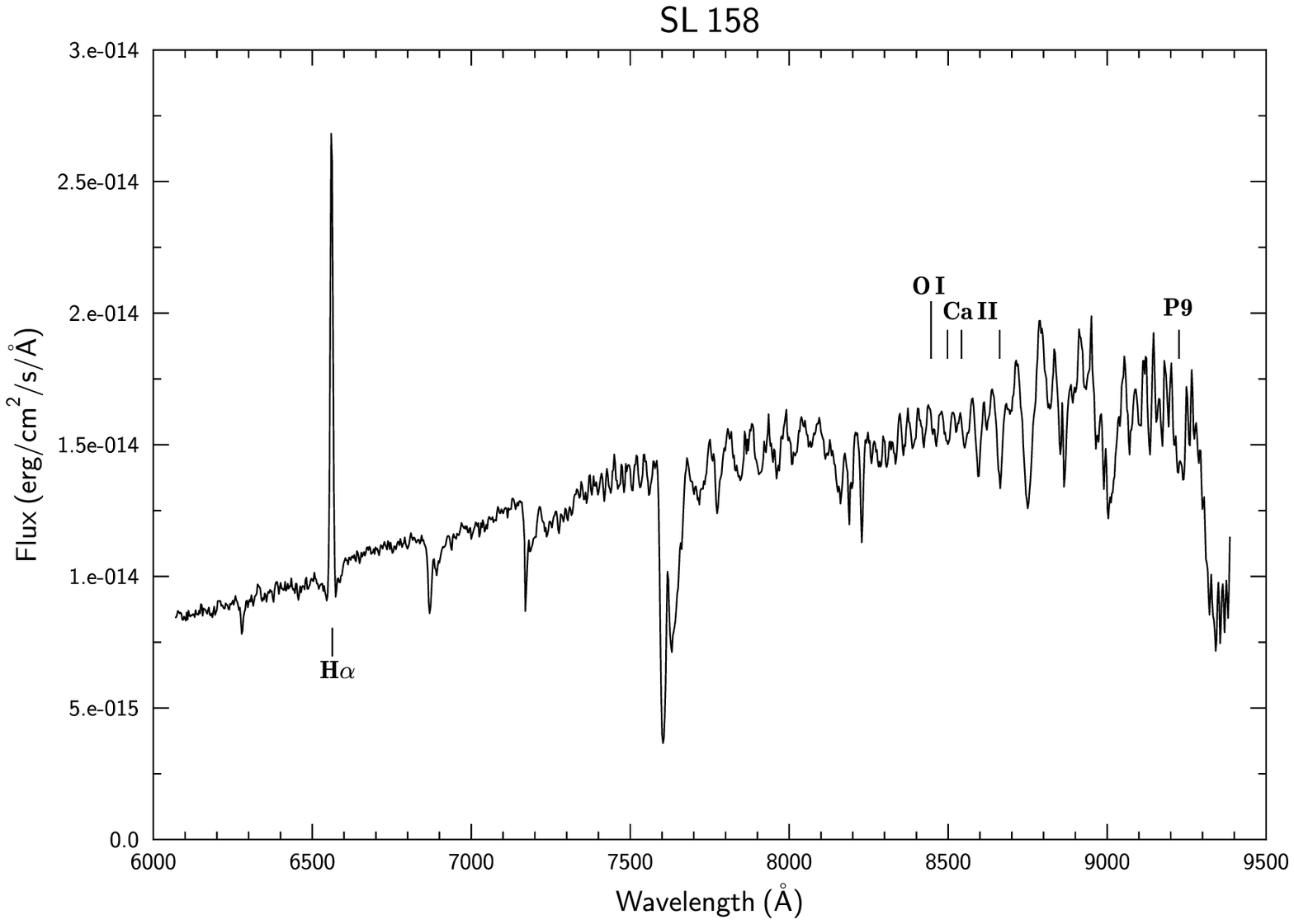,width=124mm,angle=0,clip=true}
\vskip.5mm
\captionc{4g~and~4h}{Spectral energy distributions for the stars
SL\,184 and GL\,158.}
\end{center}
}

\vbox{\begin{center}
{\psfig{figure=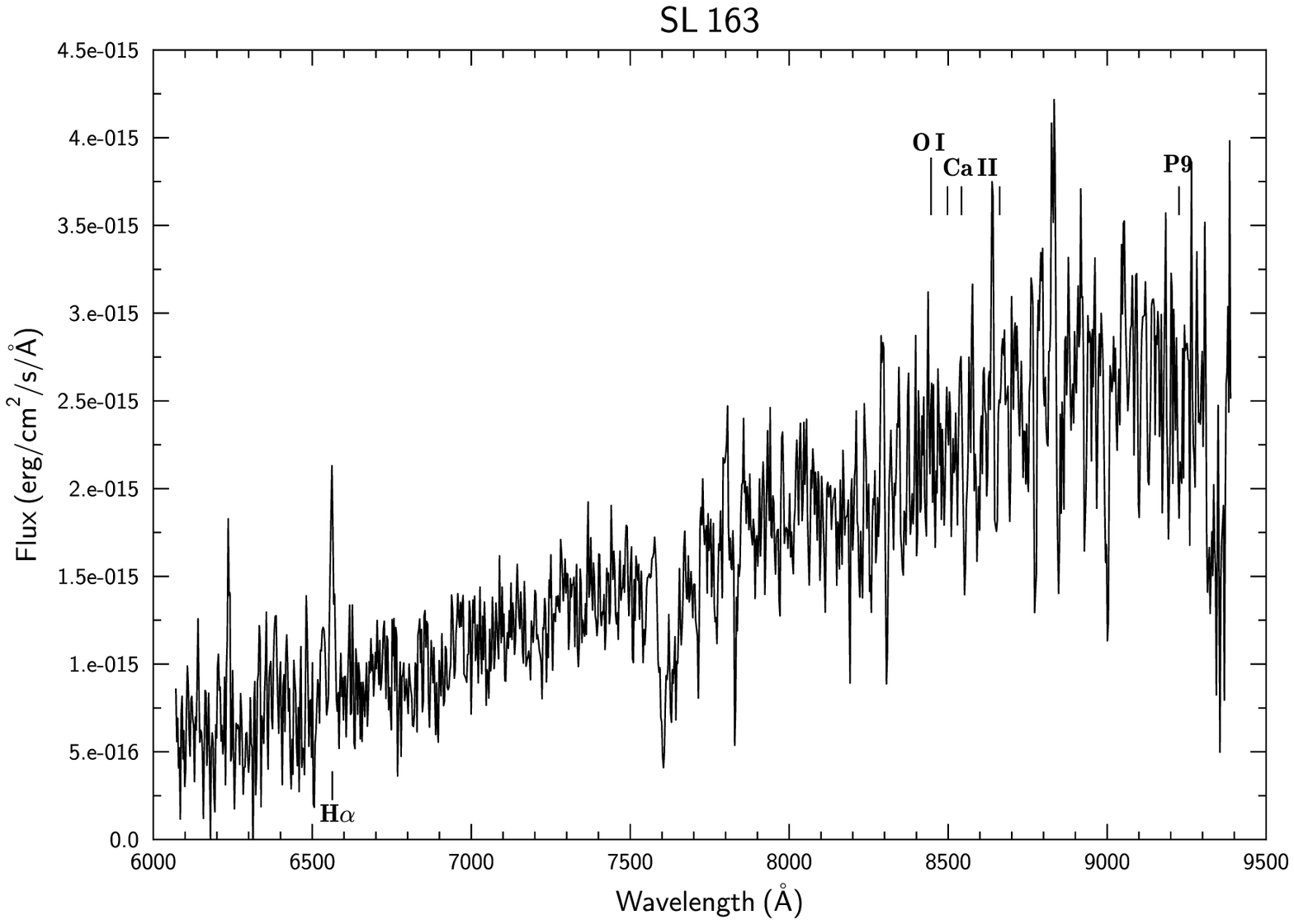,width=124mm,angle=0,clip=true}}
\vskip2mm
{\psfig{figure=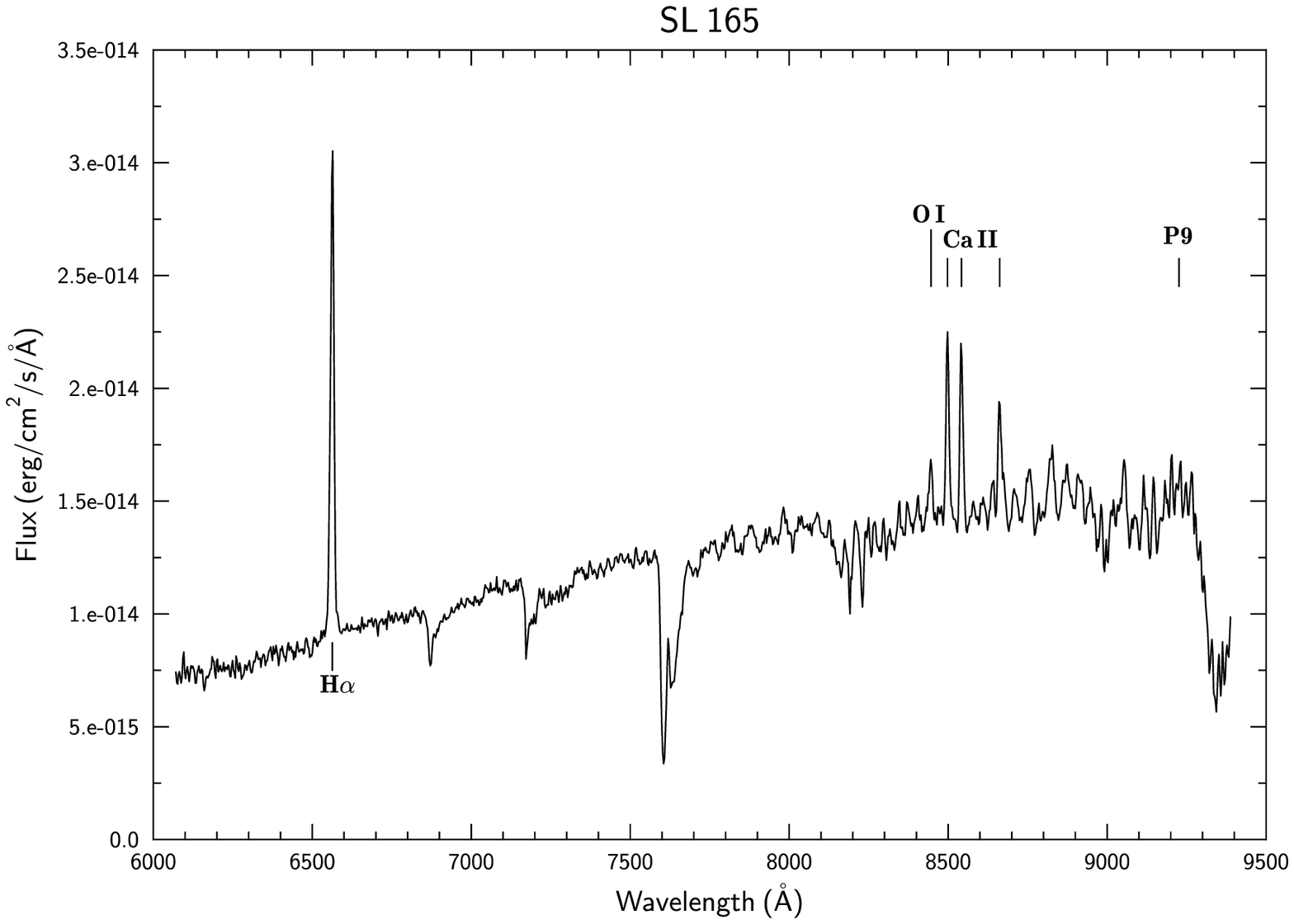,width=124mm,angle=0,clip=true}}
\vskip2mm
\captionc{4i~and~4j}{Spectral energy distributions for the stars
SL\,163 and SL\,165.}
\end{center}
}

\vbox{\begin{center}
\psfig{figure=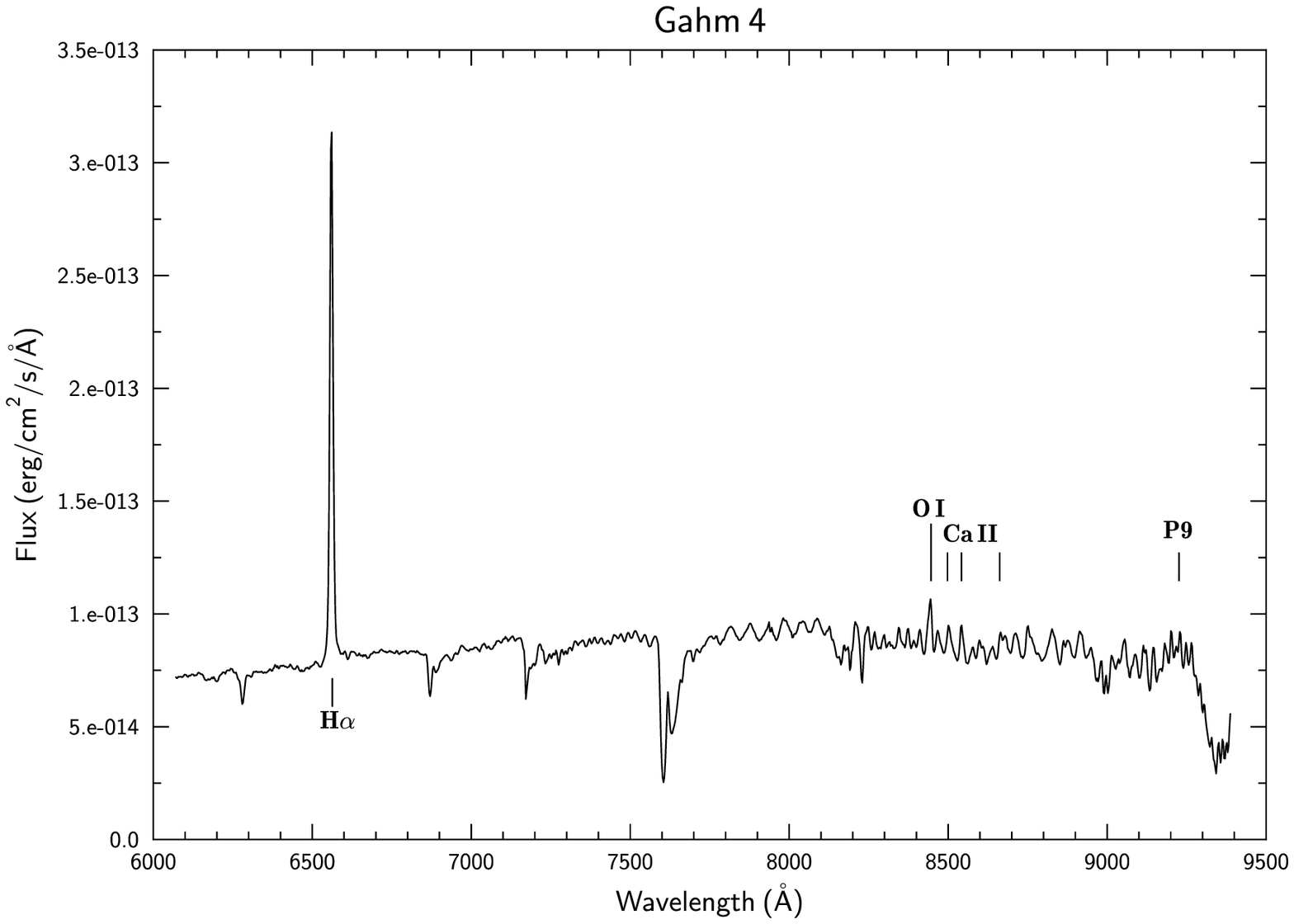,width=124mm,angle=0,clip=true}
\vskip.5mm
{\psfig{figure=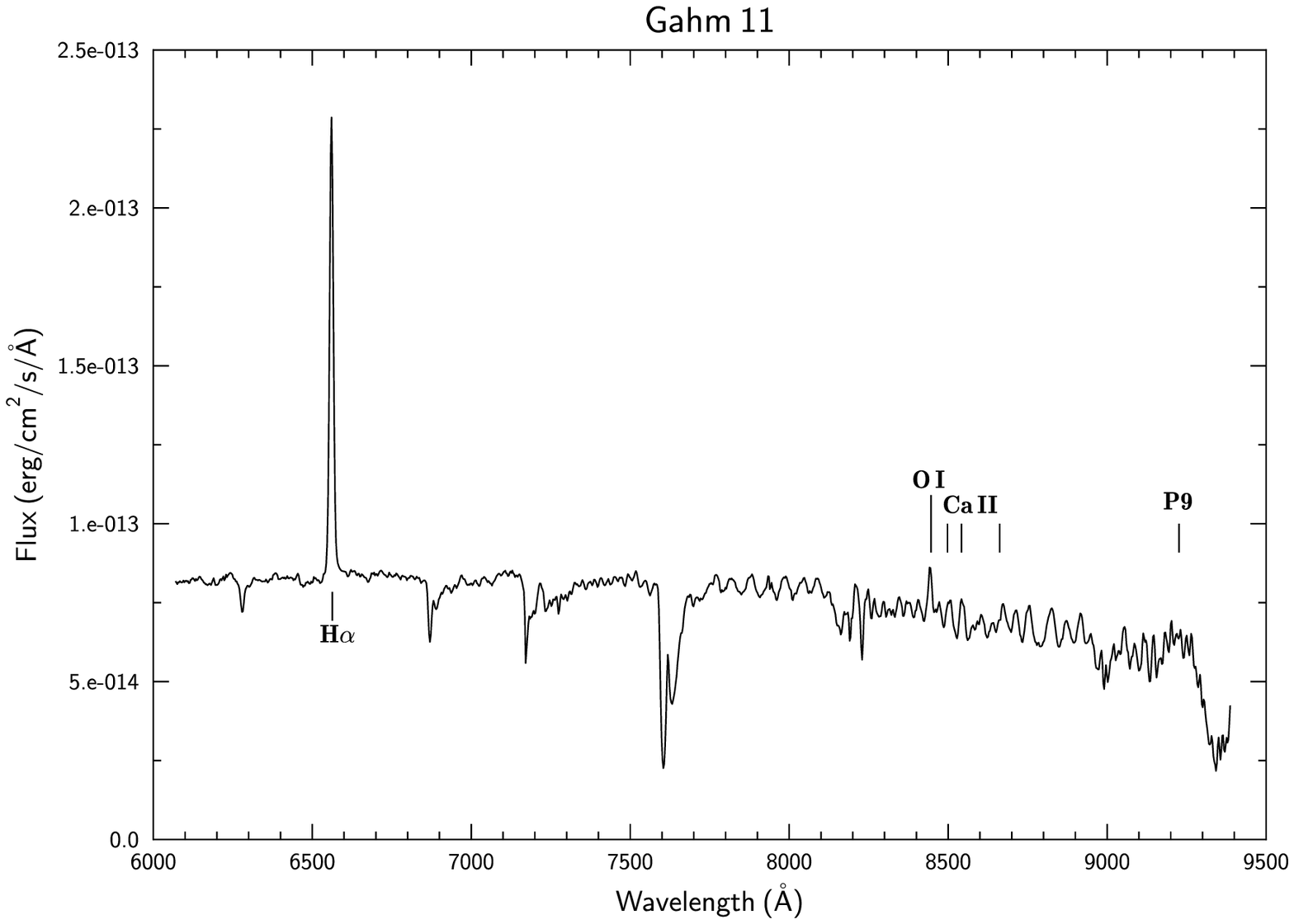,width=124mm,angle=0,clip=true}}
\vskip2mm
\captionc{4k~and~4l}{Spectral energy distributions for the stars
Gahm 4 and Gahm 11.}
\end{center}
}

\vbox{\begin{center}
\psfig{figure=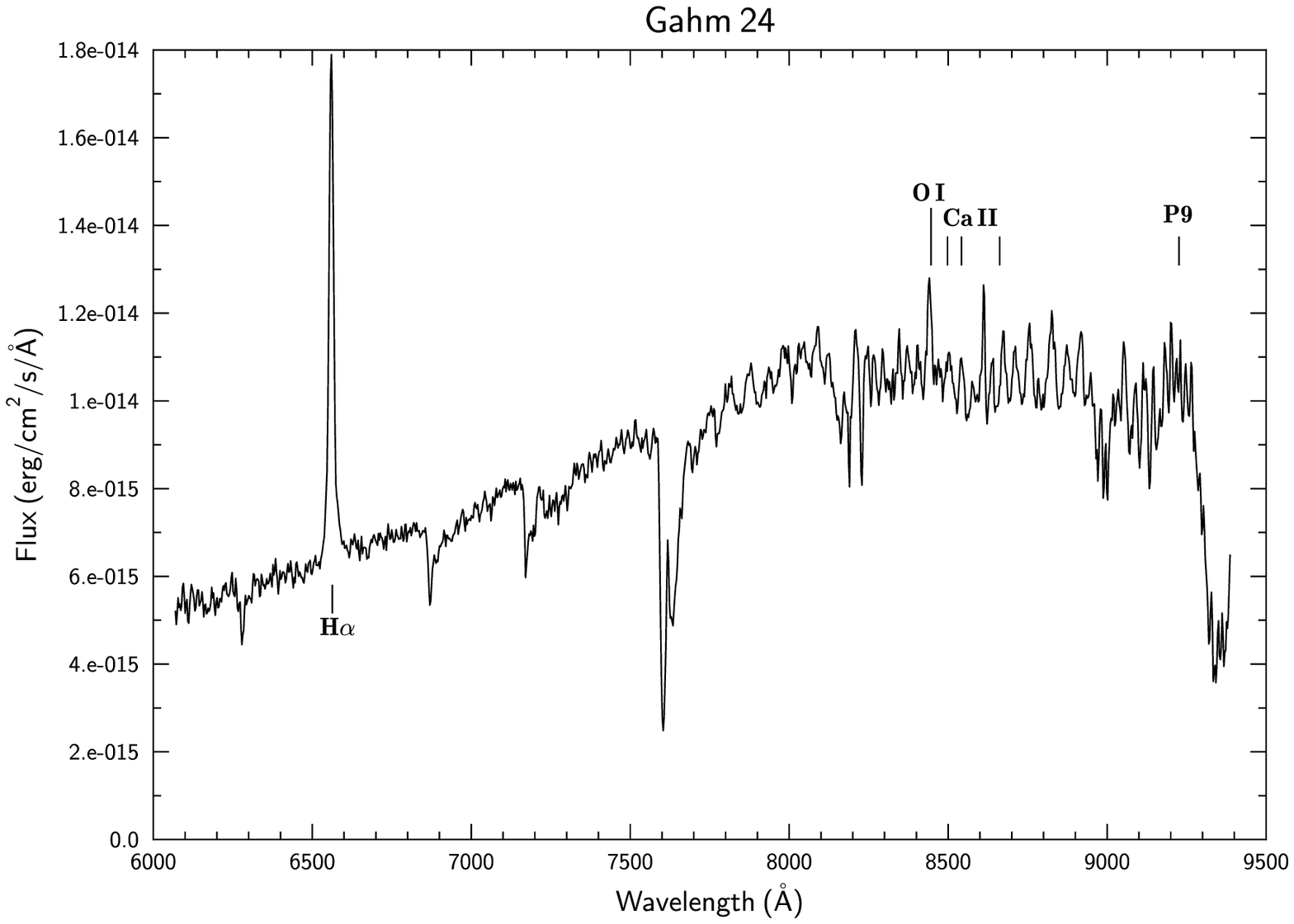,width=124mm,angle=0,clip=true}
\vskip.5mm
{\psfig{figure=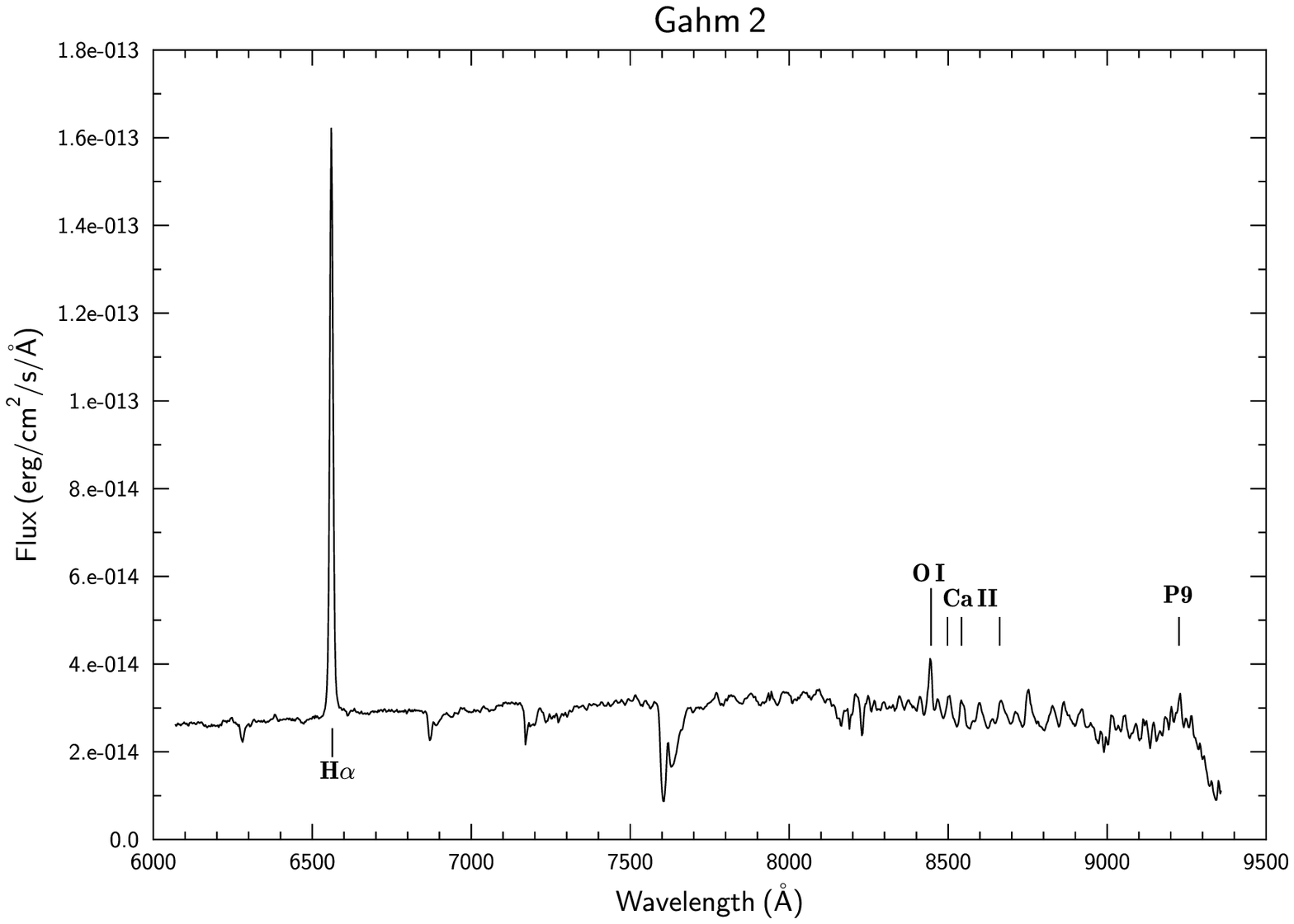,width=124mm,angle=0,clip=true}}
\vskip2mm
\captionc{4m~and~4n}{Spectral energy distributions for the stars
Gahm 24 and Gahm 2.}
\end{center}
}

\begin{figure}[!t]
\vbox{\begin{center}
\psfig{figure=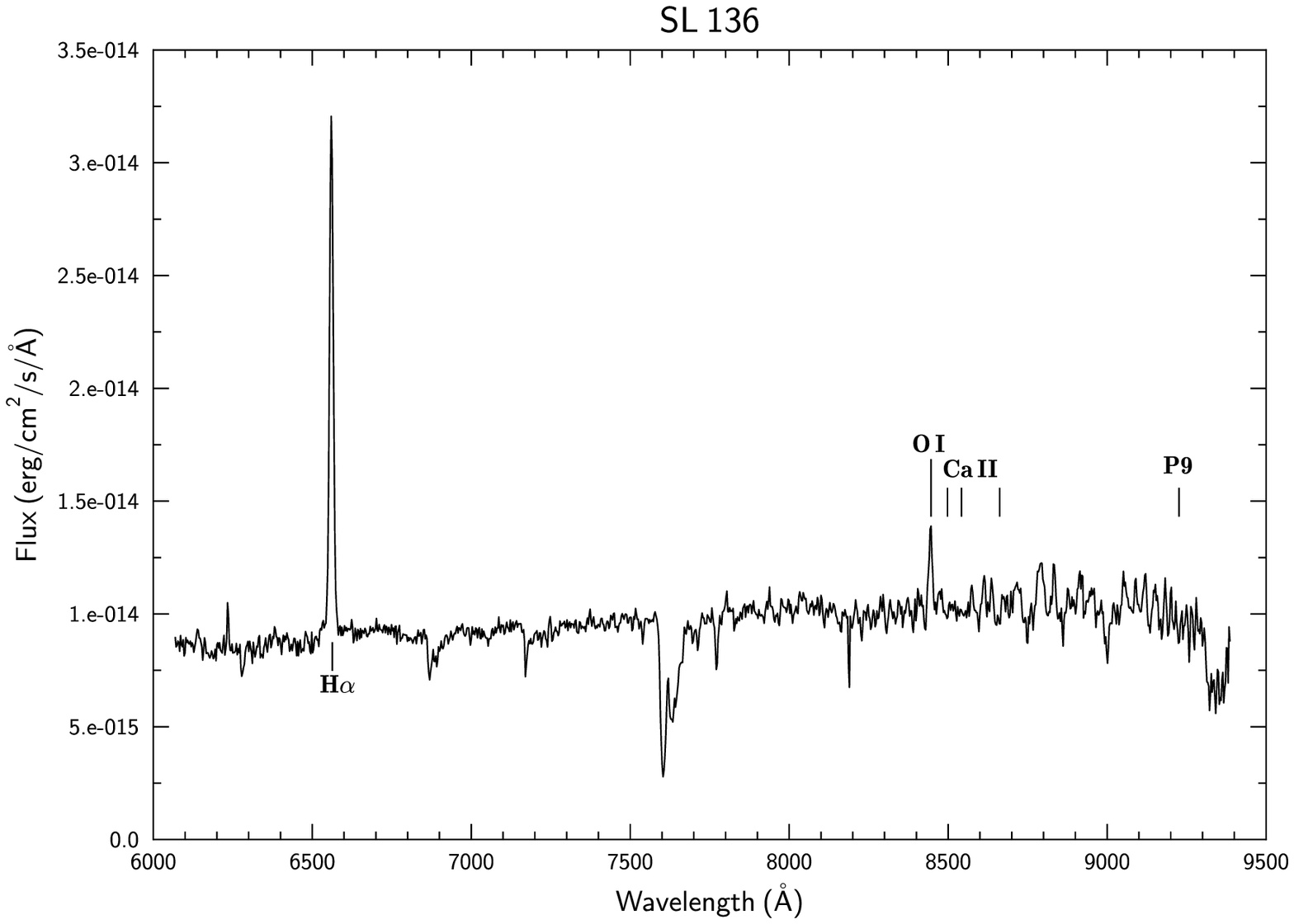,width=124mm,angle=0,clip=true}
\vspace{-8mm}
\captionc{4o}{Spectral energy distribution for the star
SL\,136.}
\end{center}
}
\end{figure}

\noindent 8.3 $\mu$m passband.  Spectral energy distribution shown in
Paper III is consistent with an YSO of class III.  Our spectrum shows
the emission in H$\alpha$ with {\it EW} = --17.1 \AA\ and spectral type
A0e.  Taking into account changes in the telluric bands, the spectrum is
quite similar to that observed in 2007.

\vskip1mm

{\bf SL\,163 = 2MASS J03325315+5827511}

This is the faintest star in our sample, its magnitude $F$ = 17.51.  Its
spectrum is very noisy, and the classification (F:\,e) uncertain.
However, the H$\alpha$ emission is quite strong ({\it EW} = 17.4 \AA).
In the $J$--$H$ vs.\,$H$--$K_s$ diagram it lies exactly on the intrinsic
T Tauri star line.  The IR source IRAS 03289+5818, located at the
89\arcsec\ angular distance from SL\,163, probably is not related.

\vskip1mm

{\bf SL\,165 = 2MASS J03340073+5816376 = MSX G142.9713+01.8792}

The object in the $J$--$H$ vs.\,$H$--$K_s$ diagram lies about 0.15 mag
above the intrinsic T Tauri star line.  The MSX flux at 8.3 $\mu$m
allows us to conclude that the star is an YSO of class II (Paper III).
Our spectrum confirms that the star is of T Tauri type with strong
emissions in H$\alpha$, O\,I and Ca\,II; its spectral type is G5e.

\vskip1mm

{\bf SL\,136 = 2MASS J04315244+4904444}

This star is located in the direction of dark cloud TGU\,1041 (Dobashi
et al. 2005) at the Galactic equator in Perseus. In the $J$--$H$
vs.\,$H$--$K_s$ diagram it lies directly on the intrinsic T Tauri star
line. The star is absent in the IRAS and MSX catalogs. Our spectrum of
the star shows strong emissions in H$\alpha$ and O\,I lines, spectral
type G0e. Most probable, it is an YSO of class II.

\newpage

{\bf The stars Gahm 4 = 2MASS J03451281+5214379, Gahm
11 = \\2MASS J03561414+5226030, Gahm 24 = 2MASS J03565522+5251200 and
Gahm 2 = 2MASS J03581379+5243109}

As was described in Paper I, Gahm (1990) discovered 12 stars with
H$\alpha$ emission in low-dispersion objective-prism spectra located
near the Sh2-205 emission nebula.  In Paper IV we confirmed the YSO
status for four Gahm stars (21, 22, 23 and 25) -- three of them were
found to fall within the temperature range of T Tauri stars and one to
be an Ae-star.  More Gahm stars were included in the present
investigation and four of them were found to show spectral properties
typical of YSOs.  All four are probably Herbig Ae stars with strong
emissions in H$\alpha$ and the O\,I line and spectral types A5e--F0e
(Table 2).  Since all these stars are absent in the IRAS and MSX
catalogs, there was no way to verify their medium infrared energy
distributions.  H$\alpha$ emission in two of the investigated stars
(Gahm 4 and 11) had been discovered earlier by Gonzalez \& Gonzalez
(1954).  These two stars as well as Gahm 2 are present in the catalog
{\it H$\alpha$ Stars in Northern Milky Way} by Kohoutek \& Wehmeyer
(1997) as KW97 16-16, 16-43 and 16-49, respectively.  Recently,
H$\alpha$ emission in Gahm 24 has been found by the IPHAS survey (Witham
et al. 2008; Gonz\'alez-Solares et al. 2008).

\vskip1mm

{\bf Stars without emission lines}

The stars in the spectra of which we did not find emission lines are
listed in Table 3. Of them three stars are in the vicinity of GL\,490
and four Gahm stars in the vicinity of Sh2-205.  The reason is not
understood since these stars in the $J$--$H$ vs.\,$H$--$K_s$ diagram lie
in the domains of T Tauri and Ae/Be stars.  Possibly, some of these
stars were misidentified in the finding charts.  For example, the star,
which we considered as Gahm 1, is located only 70\arcsec\ south of the
known emission line star KW97 16-55.  The last column of Table 3 gives
spectral classes estimated from our spectra.

\begin{table}
\begin{center}
\vbox{\small\tabcolsep=6pt
\parbox[c]{100mm}{\baselineskip=10pt
{\normbf\ \ Table 3.}{\norm\ Stars for which our spectra
do not show the presence of strong emission lines.
\lstrut}}
\begin{tabular}{lccccc}
\tablerule
 Name  &  RA\,(2000) & DEC\,(2000) &  $V$ & $F$   & Sp \\
\tablerule
SL\,180   & 3 25 19.2 &  +58 11 46 &   15.73 & 15.06 & G8 \\
SL\,171*  & 3 31 30.2 &  +57 21 56 &   18.69 & 16.60 & K3 \\
SL\,173   & 3 34 21.2 &  +56 43 43 &   14.17 & 13.63 & A5 \\
Gahm 12   & 3 49 06.0 &  +51 18 46 &   13.06 & 14.81 & A5 \\
Gahm 3    & 3 54 49.8 &  +51 13 37 &   13.98 & 13.43 & A8 \\
Gahm 1    & 4 01 17.5 &  +53 10 14 &   11.66 &       & A8 \\
Gahm 26   & 4 01 59.0 &  +53 09 44 &   11.53 &       & A8 \\
\tablerule
\end{tabular}
}
\vspace{2mm}
\parbox[c]{95mm}{\baselineskip=10pt
\hangindent12mm {\bf Note:} in SL\,171 H$\alpha$ is filled in by
emission, i.e., it is a marginal H$\alpha$ emission star.}
\end{center}
\vspace{-6mm}
\end{table}

\sectionb{4}{CONCLUSIONS}

The far-red slit spectra of 15 suspected YSOs embedded in dust and
mo\-le\-cular clouds of Camelopardalis and the nearby region of Perseus
are obtained.  Ten of the investigated stars are located in the vicinity
of the young stellar object GL\,490 in the dark cloud TGU\,942, four
objects in the vicinity of the H\,II region Sh2-205 and one object in
the dark cloud TGU\,1041 in the northern Perseus.  All of these objects
exhibit emission of different intensity in their H$\alpha$ lines and
some of them fainter emission in their O\,I and Ca\,II lines.  The
equivalent widths of the emission lines were measured and spectral
classes were estimated from the absorption features.  To assess the
evolutionary status of these stars we made use of their positions in the
$J$--$H$\,vs.\,$H$--$K_s$ diagram and spectral energy distribution
curves in the infrared determined from 2MASS {\it J, H, K}$_s$, IRAS and
MSX photometry.

In Fig.\,9 of Paper III the color-magnitude diagram $K_s$ vs.
$H$--$K_s$ was plotted for 25 YSOs located in this area, whose
pre-main-sequence status was confirmed by IRAS and/or MSX photometry.
Now we have spectroscopic confirmation of 14 of these stars (Paper IV
and this paper), all brighter than $K_s$ = 11.4.  Among these brightest
stars of the SFR we have GL\,490 (very young YSO of class I), five
A-type objects, four F-type objects and four G-type objects.  YSOs of
types K and M at a distance of 900 pc are probably fainter than the red
magnitude $F$ = 17--17.5, the limit of our spectral observations.  Among
the observed YSOs two are of class I, eight of class II and four of
class III.

The comparison of color-magnitude diagrams of the GL\,490 and Taurus
SFRs (Paper III) leads to the conclusion that YSOs in Camelopardalis are
much more massive and younger (of spectral types A to G) than in Taurus.
The GL\,490 region is also surrounded by the Cam OB1-A association
containing tens of O- and early B-stars and some cooler supergiants
(Paper I).  One of them, HD\,21389 (A0\,Ia), illuminates the reflection
nebula vdB\,15 in which the GL\,490 and other young stars are immersed.
In Taurus, YSOs of K and M types dominate, stars of spectral types F and
G are very rare, and Herbig A stars are absent.  YSOs of types F and G
are rare also in other SFRs, even in the young open cluster NGC\,2264
which contains the pre-main-sequence of a very broad range of
temperatures.  According to Cohen \& Kuhi (1979), H$\alpha$ intensities
in pre-main-sequence F--G stars vary from zero to {\it EW}\,=\,55 \AA\
(LkH$\alpha$ 338 in the Mon R2 region).  This is consistent with our
results found in the GL\,490 region.

Another group of eight confirmed YSOs, discovered by G. Gahm, is related
to the SFR close to H\,II region Sh2-205 near the Camelopardalis and
Perseus border at the Galactic longitude 149\degr.  In Paper IV and this
paper in this region we have identified YSOs of the following spectral
types:  two are A-stars, three F-stars, one G-star and two K-stars.
Recently, star formation in the region has been investigated by Romero
\& Cappa (2009) and more YSOs were suspected.

In summary, our spectral observations of suspected YSOs have increased
the number of confirmed pre-main-sequence stars in the Camelopardalis
dark clouds to 22.  The results prove that the dark cloud
surrounding a massive protostar GL\,490 and the region near the H\,II
cloud Sh2-205 are regions of active formation of stars more massive
than the Sun.

\thanks {We are thankful to the Steward Observatory for the observing
time and to Edmundas Mei\v{s}tas for his help preparing the paper.  The
use of the 2MASS, IRAS, MSX, IPHAS, SkyView, Gator and Simbad databases
and the IRAF program package is acknowledged.}

\References

\refb Cohen M., Kuhi L. V. 1979, ApJS, 41, 743

\refb Corbally C. J., Strai\v{z}ys V. 2008, Baltic Astronomy, 17, 21
(Paper IV)

\refb Danks A. C., Dennefeld M. 1994, PASP, 106, 382

\refb Dobashi K., Uehara H., Kandori R., Sakurai T., Kaiden M.,
Umemoto T.,\\ Sato F. 2005, PASJ, 57, S1

\refb Gahm G. 1990, personal communication

\refb Gonzalez G., Gonzalez G. 1954, Bol. Obs. Tonantz. Tacub., 1-9, 3

\refb Gonz\'alez-Solares E. A., Walton N. A., Greimel R., Drew J. E. et
al. 2008, MNRAS, 388, 89

\refb Herbig G. H. 1962, Advances in Astron. \& Astrophys., 1, 47

\refb Kohoutek L., Wehmeyer R. 1997, {\it H-alpha Stars in Northern
Milky Way}, Abh. Hamburger Sternw., 11, Teil 1+2 = CDS catalog III/205

\refb Lasker B. M., Lattanzi M. G., McLean B. J. et al. 2008, AJ, 136,
735 =\\ CDS catalog I/305

\refb Meyer M. R., Calvert N., Hillenbrand L. A. 1997, AJ, 114, 288

\refb Romero G. A., Cappa C. E. 2009, MNRAS, 395, 2095

\refb Strai\v zys V., Laugalys V. 2007a, Baltic Astronomy, 16, 167
(Paper I)

\refb Strai\v zys V., Laugalys V. 2007b, Baltic Astronomy, 16, 327
(Paper II)

\refb Strai\v zys V., Laugalys V. 2008a, Baltic Astronomy, 17, 1
(Paper III)

\refb Strai\v zys V., Laugalys V. 2008b, {\it Young Stars and Clouds in
Camelopardalis}, in {\it Handbook of Star Forming Regions, vol.\,1. The
Northern Sky}, ed. B. Reipurth, ASP, p.\,294

\refb Witham A. R., Knigge C., Drew J. E., Greimel R. et al. 2008,
MNRAS, 384, 1277

\end{document}